\documentclass[12pt]{article} 

\usepackage{url}
\usepackage{setspace}
\usepackage{amsmath,amssymb,amscd}
\usepackage{amsfonts}
\usepackage{color}
\usepackage{graphicx}

 \newcommand{\bea}{\begin{eqnarray}}
\newcommand{\eea}{\end{eqnarray}}
\newcommand{\be}{\begin{equation}}
\newcommand{\ee}{\end{equation}}
\newcommand{\ba}{\begin{align}}
\newcommand{\ea}{\end{align}}

\newcommand{\K}{\mathcal{K}}

\newcommand{\C}{\mathcal{C}}

\newcommand{\cH}{\mathcal{H}}

\newcommand{\Z}{\mathbb{Z}}

\newcommand{\ie}{{\it i.e.~}}
\def\eg{{\it e.g.~}}

\newlength{\slength}
\newcommand*{\bb}{\makebox[\slength][c]{$\bullet$}}
\newcommand*{\wb}{\makebox[\slength][c]{$\circ$}}
\newcommand*{\tw}{\makebox[\slength][c]{$\times$}}

\newcommand\stab[1]{G_N^{\mathcal #1}}

\newcommand\wrS{S_{\sqrt{N}}\wr S_{\sqrt{N}}}

\begin{document}

\title{\vspace{-1cm}\begin{flushright}{\small RUNHETC-2015-09}\end{flushright}\vspace{2cm}\bf{Permutation Orbifolds in the large $N$ Limit}}

\author{Alexandre Belin$^\blacklozenge$, Christoph A. Keller$^\spadesuit$, Alexander Maloney$^{\bigstar}$}
\maketitle
\begin{center}
$^\blacklozenge$\ {\it Stanford Institute for Theoretical Physics and Department of Physics,\\ \smallskip Stanford University, Stanford, CA 94305, USA} \\ \bigskip
$^{\spadesuit}$\ {\it NHETC, Rutgers, The State University of New Jersey, Piscataway, USA}\\ \bigskip
$^{\bigstar}$\ {\it  Department of Physics, McGill University, Montr\'eal, Canada } \\ \smallskip

\vspace{2em} 
\texttt{alexandrebelin1986@gmail.com, keller@physics.rutgers.edu, maloney@physics.mcgill.ca}
\end{center}
\vspace{3em}

\begin{center}
{\bf Abstract}
\end{center}
The space of permutation orbifolds is a simple landscape of two
dimensional CFTs,
generalizing the well-known symmetric orbifolds.
We consider constraints which a permutation orbifold with large
central charge must obey in order to be holographically dual to a weakly coupled (but possibly stringy) theory of gravity in
AdS. 
We then construct explicit examples of permutation orbifolds which obey these constraints.
In our constructions the spectrum remains finite at large $N$,
but differs qualitatively from that of symmetric orbifolds.
We also discuss under what conditions the correlation functions factorize
at large $N$ and thus reduce to those of a generalized free field in AdS.
We show that this happens not just for symmetric orbifolds, but also
for permutation
groups which act ``democratically" in a sense which we define.

%

\newpage

\tableofcontents

\newpage

\section{Introduction}

\subsection{AdS/CFT and the Space of CFT$_2$'s}

The AdS/CFT correspondence provides, at least in principle, a completely non-perturbative definition of quantum gravity in asymptotically Anti-de Sitter Space.  Weakly coupled theories of AdS gravity are  typically dual to strongly coupled CFTs, making it difficult to use this correspondence to make precise statements about semi-classical gravity.  The AdS$_3$/CFT$_2$ correspondence, however, provides the hope of something more.  The constraints of conformal invariance in two dimensions are much stronger than in $d>2$, allowing us to understand the duality even in the semi-classical limit.  
For example, the infinite Virasoro symmetry of two dimensional CFTs can be understood as the asymptotic symmetry algebra of three dimensional gravity  \cite{Brown:1986nw}.  Similarly, modular invariance -- i.e. invariance under large conformal transformations -- can be understood as coming from a sum over semi-classical saddle points of the gravitational path integral \cite{Dijkgraaf:2000fq, Maloney:2007ud}.  

In the AdS$_3$/CFT$_2$ dictionary, a CFT with central charge $c$ is dual to a theory of gravity with 
\be
c = {3 \ell \over 2G}\ ,
\ee
where $\ell$ is the AdS radius and $G$ is Newton's constant.  
The space of two dimensional CFTs can be regarded as a landscape of theories of quantum gravity in AdS$_3$, with many different types of matter content and values of the coupling constant $c$.  A weakly coupled theory of gravity -- that is, one where $G$ is small -- is dual to a CFT with large central charge $c$.   In this paper we will characterize a simple class of conformal field theories at large central charge and understand features of the corresponding gravity duals.

The set of two dimensional CFTs -- while much simpler than in higher dimensions -- is still quite complicated.  The best understood theories are rational CFTs, which have small values of $c$.  The Virasoro minimal models, with $c<1$, are the most well known examples.  Exactly solvable CFTs with $c>1$ can also be constructed, but they are always rational with respect to some larger chiral algebra which includes not just the Virasoro generators but also some non-trivial ${\cal W}$-symmetries.  In other words, rational CFTs may have $c>1$, but the central charge will still be ``small" compared to the size of the symmetry algebra.  Most attempts to describe pure theories of quantum gravity in AdS (i.e. theories with only metric degrees of freedom) involve rational CFTs of some type.\footnote{See e.g. \cite{Castro:2011zq, Gaberdiel:2010pz, Maloney:2009ck, Witten:2007kt} for candidate duals to minimal model CFTs.  We note that the extremal CFTs of Witten \cite{Witten:2007kt} are also rational CFTs in a rather trivial sense; they are chiral CFTs, so by definition are rational with respect to a sufficiently large chiral algebra.}  In this paper we will take the opposite approach: rather than trying to study a simple, exactly solvable CFT which is dual to a particularly simple theory of gravity, we will attempt to characterize more generally the space of conformal field theories with large central charge.

The characterization of the space of large $c$ CFTs is an interesting problem in its own right. 
Conformal bootstrap techniques have proven useful in constraining the general structure 
of 2d holographic CFTs \cite{Rattazzi:2008pe, Hellerman:2009bu,Friedan:2013cba,Qualls:2013eha,
Hartman:2014oaa,Keller:2014xba,Qualls:2014oea,Qualls:2015bta},
as well as other aspects such as locality and
thermalization \cite{Fitzpatrick:2012yx,Fitzpatrick:2014vua,Fitzpatrick:2015zha}.
Unfortunately a complete classification still seems out of reach.
We will therefore consider only a particularly tractable corner of the landscape of 2d CFTs: the space of permutation orbifolds.
The virtue of this approach
is that, at least in principle, one can construct explicitly 
all theories in this subspace. 
 The simplest examples are symmetric orbifolds, which appear as the
dual CFTs in various string theory constrictions of AdS$_3$, including  the D1-D5 system.
Indeed, this family appears to include
all known explicit examples of holographic CFTs with
large central charge.

\subsection{Holographic CFTs}

In this paper we are interested in theories of gravity with a semi-classical limit.  This means we will consider not a single CFT, but rather a family of CFTs labeled by a parameter $N$ which is proportional to the central charge.  The semiclassical limit is $N\to\infty$. 
Although any individual CFT can be interpreted as a theory of gravity in AdS -- at least in the sense that the CFT correlation functions can be regarded as scattering amplitudes for fields in asymptotically Anti-de Sitter space -- the $N\to\infty$ limit may not describe a well behaved weakly coupled theory of gravity.  We emphasize that, in the present context,  by a ``weakly-coupled" theory of gravity we do not necessarily mean that the dual gravity theory is perturbative Einstein gravity coupled to matter.  We only mean that -- since the Planck length is small in AdS units -- gravitational backreaction is negligible.
For example, we will not require out theory to be local on the AdS scale; this would require additional constraints, such as the existence of a large gap in the CFT spectrum \cite{Heemskerk:2009pn}. 
We would, for example, be happy to consider theories of gravity which have as their $N\to\infty$ limit a classical ($g_s\to0$) string theory with string length of order the AdS scale.  Such theories are expected to be dual to weakly coupled gauge theories in the large $N$ limit.

We now ask what conditions we must impose on these theories in order for a well defined semi-classical limit to exist; see \cite{Belin:2014fna, Haehl:2014yla, ElShowk:2011ag} for similar considerations.
Our first constraint is that the number of states should remain finite in the large $N$ limit.  More precisely, if we let $\rho_N(\Delta)$ be the number of states with energy (i.e. scaling dimension) $\Delta$ in theory $N$, then we demand that the limit
\be
\rho_\infty(\Delta) = \lim_{N\to\infty} \rho_N(\Delta)
\ee
exists and is finite for any $\Delta$.  
 This is the statement that, once gravitational backreaction is turned off, the theory has only a finite number of degrees of freedom below a given energy.  This property is satisfied by any of the familiar examples of AdS/CFT, including large $N$ gauge theories.   Indeed, once we take $N\to\infty$, the resulting function $\rho_{\infty}(\Delta)$ can be used to characterize the semi-classical gravity theory.  In particular, the states counted by $\rho_\infty(\Delta)$ are interpreted as perturbative excitations in AdS in the limit where  gravitational back-reaction is neglected.  In a typical string theory, for example, one would expect a Hagedorn spectrum
 \be
 \rho_\infty(\Delta) \approx \exp\left\{\beta_H\Delta \right\}~~~~~{\rm as}~\Delta\to\infty\ ,
 \ee
 where the Hagedorn temperature $\beta_H$ is related to the string tension.\footnote{In many cases, the Hagedorn divergence can be naturally interpreted as the Hawking-Page transition between the thermal and black hole phases \cite{Aharony:2003sx, Keller:2011xi}.}  In a theory with fewer degrees of freedom  $\rho_\infty(\Delta)$  would increase more slowly with $\Delta$.  In particular, for a local quantum field theory in $d$ dimensions compactified down to AdS$_3$ we expect
 \be\label{QFTgrowth}
 \rho_\infty(\Delta) \approx  \exp\left\{ \beta \Delta^{(d-1)/d}\right\} ~~~~~{\rm as}~\Delta\to\infty\ .
 \ee

It is worth emphasizing that the semi-classical density of states $\rho_\infty(\Delta)$ does not obey Cardy's formula.  In particular, the states which exhibit Cardy growth have $\Delta \gtrsim O(N)$, so are removed from the spectrum if we keep $\Delta$ fixed as $N\to\infty$.    Indeed, these states are interpreted as BTZ black holes, which have very high energy in the limit  $G\to 0$ where gravitational interactions are turned off.
  
 Of course, we expect that holographic CFTs dual to semi-classical gravity should exhibit other features in addition to a finite spectrum.  For example, we must also demand that the correlation functions remain finite in the large $N$ limit.  More precisely, we require that for any $\Delta$, the spectrum of operators with energy $<\Delta$ must stabilize for sufficiently large $N$.  Moreover, we require that the correlation functions for (correctly normalized) operators at finite separation will approach well defined, finite limits as $N\to \infty$.  
 
 Given these assumptions, we may then ask whether a given family of CFTs has other features which resemble semi-classical gravity in AdS.  For example, we can ask whether the correlation functions factorize into products of two point functions at large $N$, signifying that the bulk theory reduces to a linearized theory of generalized free fields. In this paper we will restrict our attention to a family of theories where these questions can be addressed precisely: Permutation Orbifolds.
 
\subsection{Permutation Orbifolds}

The simplest way to construct a large central charge CFT is to take $N$ non-interacting copies of a given "seed" CFT $\mathcal{C}$ with Hilbert space $\mathcal{H}$ and central charge $c$.  The resulting tensor product theory $\mathcal{C}^{\otimes N}$ has central charge $N c$. Such theories are not, however, good candidate holographic duals of semi-classical gravity as they will typically have an infinite number of states at low energies.  
To obtain a finite number of states, we will consider orbifold theories.  In particular, the product theory $\mathcal{C}^{\otimes N}$ has $S_N$ global symmetry that interchanges the various copies.  One can quotient the theory by any subgroup $G_N \subseteq S_N$. A theory obtained in this manner is called a permutation orbifold and denoted
\be
\mathcal{C}_{G_N}=\frac{\mathcal{C}^{\otimes N}}{G_N}\ .
\ee
The orbifold theory will typically have two types of states: twisted sector  and untwisted sector states. The untwisted sector is obtained by simply taking the subset of $\mathcal{H}^{\otimes N}$ that is invariant under the action of $G_N$. For the theory to have a well-defined $N\to\infty$ limit, the total density of states has to be finite for any state of finite energy. This places strong constraints on the subgroups we can consider. For example, one can easily show that the cyclic group $\mathbb{Z}_N$ does not have this property. As we will argue, subgroups with this property must have a finite number of orbits on $K$-tuples as $N\to\infty$. They are called \textit{oligomorphic} families of groups and we describe their properties in the following subsection.

The landscape of permutation orbifolds is, of course, just a tiny corner of the space of CFTs.  It is, however, a rich enough space that it describes a variety of low energy spectra and correlation functions in the semi-classical limit.
%
%
In the first part of this paper we will investigate the spectra of permutation orbifolds at large $N$.  We will establish several results for the spectrum which hold universally, and rule out the QFT growth of the type (\ref{QFTgrowth}).  This extends results found previously in \cite{Belin:2014fna, Haehl:2014yla}.
We will also describe a new class of examples whose spectra exhibit novel and interesting features that differ from the more familiar symmetric products.  
We pay particular attention to the wreath product $S_{\sqrt{N}} \wr S_{\sqrt{N}}$ and tensor product $S_{\sqrt{N}} \times S_{\sqrt{N}}$ theories, and present several explicit results for their spectra. 
In the second part of the paper we will investigate correlation functions.  We will show that symmetric product orbifold correlation functions factorize in the way expected of free field correlators, and describe the circumstances under which this feature generalizes to other permutation orbifolds at large $N$.

Ultimately, our goal is to understand the statistics of the space
of permutation orbifolds, and to describe features of ``generic" conformal field theories in the large $N$ limit, along the lines of \cite{Benjamin:2015hsa, Moore:2015bba}.  This would allow us to understand, for example, how likely
it is that a randomly chosen family of CFTs happens be dual to a theory of weakly coupled gravity in AdS.

\section{Permutation Orbifolds and their spectrum}

\subsection{Oligomorphic families $G_N$}
We will first discuss the spectrum of permutation
orbifolds. Starting out with an arbitrary seed theory
$\C$, we take its $N$-fold tensor product $\bigotimes^N \C$. Regardless
of any symmetries of $\C$, $\bigotimes^N \C$ is symmetric
under permutations $g \in S_N$. Using the standard orbifold construction 
\cite{Dijkgraaf:1989hb,Ginsparg:1988ui},
we can orbifold the theory by any permutation group $G_N \subset S_N$.
This means we project onto states $\Phi\in \mathcal{H}^{\otimes N}$
which are invariant under $G_N$.  Since the resulting theory
is no longer modular invariant, we need to add in so-called
twisted sectors to restore modular invariance. For each conjugacy class $[g]$
in $G_N$
we need to add one corresponding twisted sector. These twisted states of course also need
to be invariant under $G_N$, or more precisely under the 
centralizer $C_g$ of $g$. The net effect of an orbifold
is thus not to so much to eliminate states, but rather 
to rearrange them. Still, for our purposes this will often
be enough. The important point is that for permutation orbifolds
most (but not all) twisted states have weight $\Delta \sim cN/12$,
and are therefore harmless in the $N\rightarrow\infty$ limit.

We are interested in the spectrum $\rho_{G_N}(\Delta)$
in the $N\rightarrow\infty$ limit. Even though we will not
do so, it should nonetheless be possible 
to make the notion of a limit of families of CFTs precise.
The rough idea is to ensure that for any $\Delta_1$, both the spectrum and
all correlation functions of states with $\Delta < \Delta_1$
converge. Note that this means that we are only interested
in states with finite $\Delta$ in the large $N$ limit. 
Also note that for each $N$, the theory will have Cardy behavior
for $\Delta \gg cN$, the onset of this behavior diverges with $N$.
The actual behavior we see will therefore be quite different
from Cardy behavior, even for states whose weight $\Delta$
is much higher than the central charge $c$ of the seed theory $\C$.

Not surprisingly, analyzing the untwisted sector is much easier
than the twisted sectors. We will thus begin with the untwisted
states.

\subsubsection{Untwisted states}

Let us now consider a family of permutation groups $G_N\subseteq S_N$.
We first need to describe the states of such a theory in some detail.
To construct a generic untwisted state,
we start with a state $\phi$ in the underlying tensor theory
$\mathcal{C}^{\otimes N}$. Such a state will consist of
$K$ factors which are in some non-vacuum states $\varphi_i$
of the seed theory, while the rest of the factors will be in the vacuum.
In particular, states of the $N\to\infty$ theory will have finite weight $\Delta$ only if almost all factors are in the vacuum. 
We can thus label such a state by an ordered $K$-tuple $\vec{K}$ of
distinct integers, and a $K$-vector $\vec{\varphi}$
of states in the seed theory,
\be
\phi = \phi_{(\vec{K},\vec{\varphi})}\ .
\ee
The notation here is that the state $\varphi_i$ is in factor
$K_i$, and all factors not specified by $\vec{K}$ are in the vacuum.
An advantage of this notation is that it does not depend
explicitly on $N$. Taking the $N\rightarrow\infty$ limit
on the level of such states is thus straightforward.

In the orbifolded theory, a generic state $\phi$ that lives in the product Hilbert space $\mathcal{H}^{\otimes N}$ 
will obviously not survive the projection onto $G_N$-invariant states. 
In this context we will thus call $\phi$ a prestate and use it to build actual states of the 
$G_N$ orbifolded theory. Using the notation introduced above,
$G_N$ simply acts on $\vec{K}$ in the natural way, and does not
affect $\vec{\varphi}$. We can project on an invariant state 
by summing over the images of $G_N$,
so that an actual state $\Phi$ in the orbifold theory is given by the orbit
of $\vec{K}$,
\be
\Phi = \sum_{g\in G_N} \phi_{(g.\vec{K},\vec{\varphi})} \, .
\ee

To count the number of states, we are thus lead to
counting the number of orbits of ordered $K$-tuples.
Let us denote the number of orbits under $G_N$ of ordered $K$-tuples
of distinct elements by $F_K(G_N)$. This is indeed the number
of untwisted states coming from the states $\vec{\varphi}$
if all states in $\vec{\varphi}$ are distinct. If some of them
are the same, then $F_K(G_N)$ will overcount them. For instance,
if all of them are equal, then the number of states is given
$f_K(G_N)$, the number of orbits of (unordered) subsets of $K$
distinct numbers. For a general $\vec{\varphi}$, the number
of states will lie somewhere in between. In general we have
the relation
\be\label{fvsF}
f_K \leq F_K \leq K! f_K \ .
\ee
The detailed relation between $f_K$ and $F_K$ is in general
very complicated and depends greatly on $G_N$.

To get the total number of states of weight $\leq \Delta$,
we also need to count the number of possible $\vec{\varphi}$.
The important point here is that for any $\Delta$ we will
get a finite number of such configurations, which is independent
of $G_N$. In particular it does not depend on $N$. The whole
large $N$ behavior of the spectrum is thus determined by
the $\vec{K}$-orbits of $G_N$.

Let us now consider the large $N$ limit of a family $G_N$.
We need this family to converge in an appropriate sense,
as mentioned above. We will require that the spectrum converges, i.e. that
for any fixed $\Delta_1$ the limit $N\rightarrow\infty$ leads to a finite spectrum
of states with $\Delta < \Delta_1$.
For that to happen, we need $F_K(G_N)$ to converge to a finite
number as $N\rightarrow\infty$, which means that
$F_K(G_N)$ becomes independent of $N$,
\be
F_K(G_N)= F_K \qquad \textrm{for}\ N\ \textrm{large enough.}
\ee
A family of groups $G_N$ which has this property is called
\emph{oligomorphic}. From the remarks above, it follows
that a family of permutation orbifolds has a finite 
number of untwisted states if it comes from an oligomorphic
family of groups. From (\ref{fvsF}) it is clear that
it does not matter whether we consider $F_K$ or $f_K$ here.

\subsubsection{Twisted states}
We now turn our attention to twisted states. The situation here is
slightly more complicated, but it turns out that
the end result is the same as in the untwisted sector: oligomorphic permutation orbifolds
also lead to a finite number of twisted states.

Let us show this in more detail. The twisted sectors of the theory are roughly given by elements
$g \in G_N$. More precisely, they are given by conjugacy classes
$[g]$.
Being a permutation, $g$ can be always written as a 
product of cycles.
In the special case of $S_N$, a conjugacy class $[g]$ is given by
the number of cycles of different lengths,
\be
[g]=(1)^{N_1}(2)^{N_2}\cdots(s)^{N_s}
\ee
where $\sum_n nN_n =N$. The centralizer is then \cite{Dijkgraaf:1996xw}
\be\label{Scentralizer}
C_{[g]} = S_{N_1}\times (\mathbb{Z}_2^{N_2} \rtimes S_{N_2} )\times\cdots \times (\mathbb{Z}_s^{N_s}\rtimes S_{N_s})\ .
\ee
Here the $S_{N_n}$ permute the $N_n$ cycles of length $n$, and
the $Z_n$ act as cyclic shifts within a cycle of length $n$. 
For a general permutation group $G$, $g$ can still be written
as a product of cycles, but its conjugacy classes are no
longer in one-to-one correspondence with cycle lengths.
The centralizer of $g$ in $G$ is a subgroup of 
(\ref{Scentralizer}). Note that the only part
of this centralizer that grows with $N$
are the permutations of the single cycles, which
will form a subgroup of $S_{N_1}$. In what follows
we can thus afford to be imprecise with the other
part of the centralizer.

Within a given cycle of length $n$, the ground state has weight
\be
\Delta_n = \frac{c}{24}\left(n-\frac{1}{n}\right)\ .
\ee
It follows immediately that to have finite weight
in the large $N$ limit, almost all factors have to
be in trivial cycles with $n=1$. The situation is thus
the same as in the untwisted sector, and we define
the length $K$ of a state
as the total number of factors which are 
in a non-trivial cycle or are not in the vacuum.

Let us generalize the notation introduced above to the
twisted sector.
Denote a pre-state $\phi_g$ of length $K$ in the twisted
sector $g$ as a triple
\be
\phi_g = \phi_{(P,\vec{K},\vec{\varphi})}\ ,
\ee 
where $\vec{K}$ is again a $K$-tuple of distinct elements,
$P$ is an integer partition of $K$ which we will represent
by a vector $(\lambda_1,\ldots\lambda_n)$, and $\vec{\varphi}$ is again 
a $K$-tuple
of states $\varphi_i$ of the seed theory. In particular
$\vec{K}$ again describes the positions of the non-trivial factors. 
The new datum $P$ describes the cycles of the permutation element $g$, 
\ie it determines which twisted
sector the state is in. 
More precisely, the permutation $g$ is given by the cycle decomposition
\be
g = \prod_{i=1}^n (K_{\mu_i+1}, K_{\mu_i+2},  \cdots, K_{\mu_i+\lambda_i})\prod_{k\notin \vec{K}}(k)
\ee
where we defined $\mu_i := \sum_{j=1}^{i-1}\lambda_j$.
The $\lambda_i$ thus encode the length of the cycles in $g$, and
we fill up $g$ with single cycles.


An actual state $\Phi$ is again given by an orbit of $\phi_g$
under conjugation by $G_N$, 
\be\label{twistedstate}
\Phi = A_\Phi^{-1/2}\sum_{h\in G_N} \phi_{(P,h.\vec{K},\vec{\varphi})} \ .
\ee
Here we used the fact that with our notation, 
conjugation by $h$ is the same as the natural action on $\vec{K}$.

Let us explain how (\ref{twistedstate}) compares to the
usual way of describing twisted states.
Usually, a twisted sector is given by the conjugacy
class $[g]$ of $g$.
In (\ref{twistedstate}) this is achieved
by summing over all elements of $[g]$ by
\be\label{twistconj}
\phi_{[g]} \sim \sum_{G_N} \phi_{hgh^{-1}} \sim \sum_{G_N/C_g} \phi_{hgh^{-1}}\ 
\ee
where the last two expressions differ by some overall factor.
Here we have used that $g$ is invariant under conjugation
by its centralizer $C_g$. We could thus reduce the sum
to the coset $G_N/C_g$. 
The state (\ref{twistedstate}) however is not invariant under $G_N$. To achieve this, we
need to sum $\phi_{hgh^{-1}}$ over the centralizer $C_{hgh^{-1}} = hC_g h^{-1}$.
Note that this does of course not effect $g$ as an element of $G_N$,
since $(P,h.\vec{K})$ corresponds to the same permutation as $(P,\vec{K})$
for $h\in C_g$. It does however affect the state $\phi_{(P,\vec{K},\vec{\varphi})}$,
since it changes how the states $\varphi_i$ are assigned to the different factors.
This is exactly how summing over the centralizer makes $\phi_g$ invariant.
In total we thus get
\be
\Phi\sim\sum_{h\in G_N/C_g} \sum_{hC_gh^{-1}} \phi_{hgh^{-1}}\ ,
\ee
which is indeed the same as (\ref{twistedstate}).

This establishes the desired result for the orbits of $G_N$.
We have to be more careful when counting the actual states,
since not every $\vec{\varphi}$ will give a state. In fact,
$\vec{\varphi}$ is no longer a $K$-vector, but rather an
$n$-vector instead. We assign a seed theory state to each
cycle rather than each individual factor, since the Hilbert
space $\cH_{(n)}$ of states in cycles of length $n$
is a subspace of the seed theory Hilbert space $\cH$ \cite{Dijkgraaf:1998zd},
\be
\cH_{(n)} \subset \cH\ .
\ee
If the centralizer is non-trivial, then $\cH_{(n)}$ is a proper
subspace. This means that even with our new definition of $\vec{\varphi}$
we overcount the number of states, since not every $\phi_i \in \cH$ leads
to a $C_g$ invariant state of the cycle. For our present purposes
this does not matter, since the overcounting is as always
independent of $N$.

This shows that oligomorphic permutation orbifolds also
have a finite number of twisted states.

\subsection{Oligomorphic groups}
Let us point out that oligomorphic \emph{groups} have been studied by mathematicians
\cite{MR1066691}. 
They are related to what we have defined above as
oligomorphic \emph{families}.
We start out with a permutation group $G$ of
an infinite countable set $\Omega$, say the natural numbers.
$G$ is then said to be oligomorphic if for all $K$,
it has only a finite number of orbits on $\Omega^K$,
the set of $K$-tuples of elements of $\Omega$.

At least morally speaking the limit of an oligomorphic
family $G_N$ should always give an oligomorphic group $G=G_\infty$.
The converse is much less obvious.
Given such an oligomorphic $G$, it is not clear
in general how to construct an oligomorphic family
$G_N$ whose orbifolds converge to $G$. One construction
is the following: Let $G\{N\}$ be the setwise stabilizer
of the set of the first $N$ elements, \ie the subgroup
that leaves that set invariant. Let $G(N)$ be the pointwise
stabilizer of the first $N$ elements, \ie the subgroup
that leaves each of the first $N$ elements invariant.
We can then define the quotient
\be
G_N := G\{N\}/G(N)\ ,
\ee
which gives a well-defined permutation group on the first
$N$ elements.
For the symmetric group $S_\infty$ this gives the desired
answer: $G_N$ is indeed exactly $S_N$. For a general
oligomorphic group however, this construction does not even guarantee
an oligomorphic family, as the following example illustrates.

Take the group $A=Aut(\mathbb{Q}, <)$ of order-preserving
permutations of $\mathbb{Q}$. This group for instance contains
continuous, piecewise linear maps of strictly positive rational slope
which are non-smooth only at rational points. One can show that this
group is oligomorphic, and has indeed $f_K(A)=1$ and $F_K(A)=K!$.
The above construction however gives $A_N=1$, since the permutation
that preserves the order of a finite number of elements
is the identity \cite{MR0401885}. 
The family $A_N$ is then clearly not oligomorphic
in our sense. We do not know if there is a way to construct
an oligomorphic $A_N$ whose orbifolds converge to $A$ in
an appropriate sense.

There is of course the question whether oligomorphic
groups per say have a physical interpretation, for instance
as the holographic dual to gravitational theories in $AdS$ with strictly infinite
radius, or at least as a tool to compute the leading
order terms in a $1/N$ expansion. In any case they
should tell us interesting facts about what limits
oligomorphic families can attain. 

There are in fact several interesting theorems about
the growth of $f_K$ for oligomorphic groups \cite{MR2581750}. 
It turns out that $f_K$ grows either polynomially
in $K$, or faster than
\be
f_K > \exp(K^{1/2-\epsilon})\ .
\ee
It is not clear what this gap signifies physically.
There are also examples of oligomorphic groups
that have super-Hagedorn growth. The automorphism
group of the random graph for instance has
\be
f_K \sim \exp(c K^2)\ .
\ee
Again, it is not clear if this can be written as
the limit of a oligomorphic family.

\subsection{Examples of oligomorphic families}
Let us now turn back to oligomorphic families.
One way to construct such families is the wreath product $A \wr B$
between two permutation groups $A$ and $B$.
From a physicist's point of view, orbifolding by $A \wr B$
simply means we take the permutation orbifold $B$ as the
new seed theory, and then perform the permutation group $A$.
We can for instance define $G_N = S_{\sqrt{N}}\wr S_{\sqrt{N}}$,
which corresponds to an iterated symmetric orbifold of $\sqrt{N}$. 
An alternate and more standard way of describing the action of $S_{\sqrt{N}}\wr S_{\sqrt{N}}$ 
is the following: arrange the $N$ factors into
an $\sqrt{N}\times\sqrt{N}$ matrix $T_{ij}$. The $i$th symmetric group
$S^i_{\sqrt{N}}$ acts on the elements of the $i$-th row
as $T_{ij} \to T_{i\sigma(j)}$, and the overall $S_{\sqrt{N}}$ permutes
the rows. The same construction gives the group theoretic
definition of a general wreath product $A\wr B$.

The group $S_{\sqrt{N}}\wr S_{\sqrt{N}}$ 
is oligomorphic, as can be seen from the
following argument. Take a $K$-tuple, that is pick $T$
with $K$ non-vanishing entries. We can now use the various
symmetric groups to move all non-vanishing entries to
the first $K$ columns, and then also to the first $K$ rows.
This shows that there can be at most $\binom{K^2}{K}$
orbits. Note that this group is much smaller than $S_N$,
since
\be
|S_{\sqrt{N}}\wr S_{\sqrt{N}}| = |S_{\sqrt{N}}|^{\sqrt{N}+1} \sim N^{(N+\sqrt{N})/2}\ .
\ee
This suggest another construction. Arrange again
the factors into a $\sqrt{N}\times\sqrt{N}$ matrix $T_{ij}$.
However now act with just a single $S_{\sqrt{N}}$
on all the columns, and with another  $S_{\sqrt{N}}$
on the rows, permuting the rows and columns
as $T_{\sigma^1(i)\sigma^2(j)}$, so that
both symmetric groups commute, giving a direct
product $S_{\sqrt{N}}\times S_{\sqrt{N}}$ . 
Note that even though we can write it as a direct
product of two symmetric groups, the action on
$N$ elements is very different from the standard
action of those two groups.
The same argument
as before shows again that there are at most $\binom{K^2}{K}$
orbits of $K$-tuples, \ie the group is again oligomorphic.
This group is even smaller, having
\be
|S_{\sqrt{N}}\times S_{\sqrt{N}}| \sim N^{\sqrt{N}}\ .
\ee
We can generalize this even further by arranging
the factors in a rank $d$ tensor $T_{i_1i_2\ldots i_d}$,
and acting with the direct product
$S_{N^{1/d}}\times\cdots\times S_{N^{1/d}}$ as 
$T_{\sigma^1(i_1)\sigma^2(i_2)\ldots \sigma^d(i_d)}$.
The same argument as above shows that this group is again
oligomorphic.
The size of this group is
\be
|S_{N^{1/d}}\times\cdots\times S_{N^{1/d}}| \sim N^{N^{1/d}}\ .
\ee
This suggests that the `faster than polynomial growth' criterion given in \cite{Belin:2014fna} is close to optimal.

\subsection{Growth in the untwisted sector}
\subsubsection{General considerations}
Having established that $\rho_\infty(\Delta)$
exists and is finite, we now want to investigate its
growth. In the cases we discuss, to leading
order the result turns out to be universal,
\ie almost independent of the choice of
the seed theory $\C$, with only its central
charge $c$ entering sometimes. This may seem
a bit surprising, so let us stress that this
statement really only holds to leading order.
More precisely, for $N$ large we investigate
the regime
\be
\rho_N(\Delta) \qquad \textrm{for} \qquad c \ll \Delta \ll cN \ .
\ee
In the cases we consider, 
the main contribution to $\rho_N(\Delta)$ in this regime
comes from states in the seed theory with $\Delta \gg c$,
which are well in the Cardy regime. To leading order,
their multiplicities are thus universally fixed by $c$.

In principle there are closed expressions for the
partition function of any permutation orbifold \cite{Bantay:1997ek}.
Unfortunately in practice it is technically hard to
extract the spectrum in the large $N$ limit.
It is particularly difficult to get a handle on
the twisted sectors. We will briefly return to this
in section~\ref{ss:twistedgrowth}. For the moment
let us concentrate on the untwisted sector, which
at least gives a lower bound on the total number of
states. To count these states, we will use the 
notation introduced in the previous section. 
Unfortunately this time we need to keep track of $N$-independent
combinatorial factors, and in particular understand
how $F_K$ and $f_K$ grow with $K$.

Let us first start
with the simplest case. Let $\varphi_1$ be the
lowest state of the seed theory of weight $\Delta_1$.
The configuration $\vec\varphi = (\varphi_1,\ldots,\varphi_1)$
then of course gives states $\Phi$ of weight $K\Delta_1$.
From these states alone the theory 
then has at least $f_K$ states of weight $K\Delta_1$.
In particular if $f_K$ grows faster than exponential,
we find that the theory has super-Hagedorn growth
already from the untwisted sector.

Next say we get to choose the $K$ states
in $\vec{\varphi}$ out of a total of $M$ states
in the seed theory.
Provided $M\geq K$, there are $M!/(M-K)!$ configurations
$\vec{\varphi}$ with distinct $\varphi_i$, each of which
contributes $F_K$ orbits. To obtain the total number of
states, we are however overcounting by a factor of
$K!$ since different permutations of the
entries in $\vec\varphi$ give the same states.
The total number of states with different individual
factors is thus
\be \label{numstates}
F_K \binom{ M}{K }\ .
\ee
If we want to keep track of states where some $\varphi_i$
are the same, the combinatorics become more difficult.
If $M \gg K$ however, this almost never happens, so that
we can neglect this effect. In that case (\ref{numstates}) becomes 
\be
F_K \frac{M^K}{K!}\ .
\ee
We stress again that this only applies if $M\gg K$. This means
that we cannot simply choose $K$ as big as we want while
keeping $M$ fixed. If we want to pick a large $K$,
we need to include states with large $\Delta$ to get 
a big enough $M$.

\subsubsection{$S_N$ redux}
As a warm up let us apply this to the symmetric orbifold.
In the process we will rederive the growth behavior of the symmetric orbifold
in the untwisted sector obtained in \cite{Belin:2014fna}. 
For the symmetric groups we have
\be
f_K = F_K =1 \ ,
\ee
so each $K$-tuple has exactly one orbit. Let us first work out
the contribution of the $K$-tuple states to states of
weight $\Delta$. For convenience set $c=3/(4\pi^2)$.
We will also assume that we are always in the
Cardy regime, an assumption whose consistency we will check
in the end.
The contribution from 1-tuples is then simply $e^{\sqrt{\Delta}}$.
The contribution from 2-tuples is
\be
\frac{1}{2!}\int d\delta e^{\sqrt{\delta}}e^{\sqrt{\Delta-\delta}} \sim  e^{\sqrt{2\Delta}}
\ee
where we have done a saddle point approximation, and the combinatorial
prefactor takes care of the overcounting as described in (\ref{numstates}).
Note that here we have assumed that almost all states
are distinct, since otherwise the combinatorial factor would change. This
is true, since the only case where states are the same are 
$e^{\sqrt{\Delta/2}}$ states of weight $\Delta/2$, where the
number of states is indeed much bigger than 2 if $\Delta$ is large enough.
Similarly, a $K$-tuple contributes with
\be\label{Kcontr}
\frac{1}{K!}e^{\sqrt{K\Delta}} \sim e^{\sqrt{K\Delta}- K \log K +K}\ .
\ee
Again most states are distinct as long as there are many more states
of weight $\Delta/K$ than $K$, \ie as long as
\be\label{distinct}
K \ll e^{\sqrt{\Delta/K}} \ .
\ee
For a fixed $\Delta$, we can thus maximize (\ref{Kcontr})
over $K$ to find where the maximal contribution to $\rho_\infty(\Delta)$ comes from.
We find that it comes from tuples of
length
\be\label{KDelta}
K \sim \frac{\Delta/4}{(\log \Delta/4)^2}
\ee
and gives indeed the result obtained in \cite{Belin:2014fna}
\be
\exp\left( \frac{\Delta/4}{\log \Delta/4} \right)\ .
\ee
Note that from (\ref{KDelta}), $\Delta/K \rightarrow \infty$ for large $\Delta$, 
so that both (\ref{distinct}) and using the Cardy formula are consistent.

\subsubsection{$S_{\sqrt{N}}\wr S_{\sqrt{N}}$}
Let us now turn to the wreath product $S_{\sqrt{N}}\wr S_{\sqrt{N}}$.
Here we have $f_K = p_K$, 
the number of integer partitions of $K$. To see this,
note that we can always use the permutations within the rows to move all
non-trivial entries of the matrix all the way to the left,
and then use the row permutation to order them in decreasing
number, giving a Young diagram.

Of more interest is $F_K$. Note that this time the non-trivial
entries are numbered. Using the same steps as above, $F_K$
is thus given by the number of different ways we can split $K$ distinct elements into 
different sets. Those are given by the Bell numbers $B_K$.
Their asymptotic behavior is given by \cite{MR671583}
\be
B_K \sim \exp( K\log K - K\log\log K -K)\ ,
\ee
\ie they grow slightly slower than factorially.
Plugging this into (\ref{numstates}), and doing the saddle point approximation
we obtain for the contribution of $K$-tuples
\be
e^{\sqrt{K\Delta}- K \log \log K}\ .
\ee
The maximum contribution thus comes from states of length
$K \sim \frac{\Delta/4}{(\log\log \Delta/4)^2}$ and gives a growth
of the form
\be\label{wreathUntwisted}
\exp\left( \frac{\Delta/4}{\log\log\Delta/4}\right)\ .
\ee 
The untwisted sector thus again has sub-Hagedorn growth.
However, not surprisingly the growth is parametrically faster
than for the symmetric orbifold.

\subsubsection{$S_{\sqrt{N}}\times S_{\sqrt{N}}$}
\begin{figure}[htbp]
\begin{center}
\includegraphics[height=.3\textwidth]{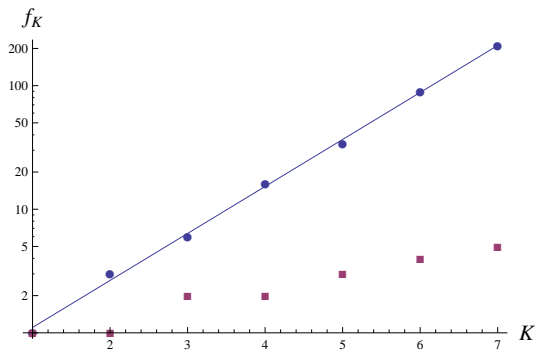}
\caption{The blue dots are the $f_K$ for 
$S_{\sqrt{N}}\times S_{\sqrt{N}}$. The blue line gives an exponential fit. For reference,
the purple squares are the $f_K$ for $S_{\sqrt{N}}\wr S_{\sqrt{N}}$.}
\label{Fig:directProduct}
\end{center}
\end{figure}

For the direct product $S_{\sqrt{N}}\times S_{\sqrt{N}}$, we have
obtained the first few $f_K$ numerically:
\be
f_K = 1,3,6,16,34,90,211\ldots
\ee
We have plotted them in figure~\ref{Fig:directProduct}. 
They grow much faster than for the wreath product, and seems to fit an exponential quite well.
Since we did not push our numerical computations very far, 
it is course possible that there are logarithmic corrections that 
make the growth slightly sub-exponential. 
If we assume that this is not the case, we can fit
\be
f_K \sim e^{\alpha K} \qquad \alpha = 0.88\ldots\ .
\ee
If the lightest non-vacuum state of the seed theory has
weight $\Delta_1$, then the number of untwisted states
of weight $\Delta$ grows at least as fast as
\be
\rho_u(\Delta) > e^{\alpha\Delta/\Delta_1}\ ,
\ee
\ie there is a Hagedorn transition already in the
untwisted sector. The Hagedorn temperature here
seems to depend on the seed theory, namely on the
weight $\Delta_1$ of its
lightest field. Note that this is only a lower bound
for the growth of total number of 
states. It is possible (indeed probable, in our view)
that the twisted states will show a super-Hagedorn behavior.

\subsection{Growth in the twisted sector}\label{ss:twistedgrowth}
Let us finally discuss twisted states. Estimating their growth
is much more involved than for the untwisted states.
Nonetheless, they are crucial for understanding the growth
behavior of all states, since in the examples we know, they
rather than the untwisted states
tend to give the dominant contribution.
In particular for symmetric orbifolds they give a Hagedorn
growth \cite{Keller:2011xi}
\be
\rho(\Delta) \sim e^{2\pi\Delta}\ .
\ee
Using the convention that
\be
Z(\beta) = \sum_\Delta \rho(\Delta) e^{-\beta(\Delta-c/24)}\ ,
\ee
the starting point for general permutation
orbifolds is Bantay's formula \cite{Bantay:1997ek}:\footnote{To simplify notation, we write the partition function $Z(\tau, {\bar \tau})$ simply as $Z(\tau)$; despite this notation, the partition function is not assumed to be a holomorphic function of $\tau$.}
\be\label{Bantay}
Z_{G_N}(\tau) = \frac{1}{|G_N|}\sum_{hg=gh} \prod_{\xi\in O(g,h)}Z(\tau_\xi)\ .
\ee
The sum here is over all $g,h\in G_N$ which commute. 
Such a commuting pair $g,h$ generates an Abelian subgroup of $S_N$,
which or course has the natural permutation action on the integers $1,2,\ldots N$.
In equation (\ref{Bantay}) we have denoted by $O(g,h)$ the set of orbits of this action. For each orbit $\xi \in O(g,h)$
we define the modified modulus $\tau_\xi$ as follows.
First, let $\lambda_\xi$ be the size of the $g$
orbit in $\xi$, and $\mu_\xi$ the number of $g$ orbits in $\xi$,
so that $\lambda_\xi \mu_\xi =|\xi|$.
Let $\kappa_\xi$ to be the smallest non-negative integer
such that $h^{\mu_\xi} g^{-\kappa_\xi}$ is in
the stabilizer of $\xi$.
Then 
\be
\tau_\xi = \frac{\mu_\xi\tau + \kappa_\xi}{\lambda_\xi}\ .
\ee
To write (\ref{Bantay}) in a maybe more familiar way,
consider a fixed $g$. This fixes a twisted sector, 
and the sum over $h$ is then a sum over the centralizer
$C_g$ which projects onto the $G_N$ invariant states in that twisted sector.
Using $|G_N|=|C_g||[g]|$, we can rewrite (\ref{Bantay}) as
as sum over conjugacy classes
\be
Z_{G_N}(\tau) = \sum_{[g]} \frac{1}{|C_g|} \sum_{h\in C_g} \prod_{\xi\in O(g,h)}Z(\tau_\xi)\ .
\ee
Let us consider a state given by a fixed $g$ of finite length $K$.
The centralizer $C_g$ of $g$ in $G_N$  is of course a subgroup
of the centralizer of $g$ in $S_N$. We can therefore write
\be
C_g = C_g^{(1)} \times C_g^{(2)}\quad \textrm{with}\quad  C_g^{(1)} \subset S_{N-K} \ , \ 
C_g^{(2)} \subset (\mathbb{Z}_2^{N_2} \rtimes S_{N_2} )\times\cdots \times (\mathbb{Z}_s^{N_s}\rtimes S_{N_s})\ .
\ee
Note that $C_g^{(2)}$ is independent of $N$.
We thus have
\be\label{Bantay2}
\frac{1}{|C_g|}\sum_{h\in C_g} \prod_{\xi\in O(g,h)}Z(\tau_\xi) = 
\left(\frac{1}{|C_g^{(1)}|}\sum_{h_1 \in C_g^{(1)}} \prod_{\xi\in O(h_1)}Z(|\xi|\tau)\right)
\left(\frac{1}{|C_g^{(2)}|}\sum_{h_2 \in C_g^{(2)}} \prod_{\xi\in O(g,h_2)}Z(\tau_\xi)\right)\ .
\ee
The first factor is simply the Polya enumeration formula,
\ie it computes the untwisted sector of a $C_g^{(1)}$ permutation
orbifold on $N-K$ factors. If $G_N$ is oligomorphic,
then $C_g^{(1)}$ is also oligomorphic. To see this,
consider orbits under $G_N$ of $K+H$ tuples $(\vec{H},\vec{K})$,
where $\vec{K}$ is the $K$-vector of all the factors in $g$.
Since $G_N$ is oligomorphic, 
we know that there are at most $F_{K+H}$ such tuples which
cannot be related by an element $a\in G_N$. Note that since
$a$ leaves $\vec{K}$ invariant, it is automatically of the form 
$C^{(1)}_g\times 1$. But this shows that $G^{(1)}_g$ as a group 
acting on $N-K$ factors has at most $F_{H+K}$ orbits of $H$-tuples,
which is independent of $N$, so that $C^{(1)}_g$ is indeed oligomorphic.
From the arguments in the untwisted sector and the fact that
the second factor in (\ref{Bantay2}) is independent of $N$,
it follows that in a given twisted sector the number of states
remains finite. Since for a given weight $\Delta$, for an oligomorphic
group $G_N$ there are only a finite number of twisted sectors that
contribute, we have reestablished the original result that oligomorphic
permutation orbifolds have indeed a finite number of states, even
when including twisted states.

Expression (\ref{Bantay2}) has a rather suggestive form.
We will try to argue that the second factor grows at most
as $e^{2\pi \Delta}$, that is in the same way as for
the symmetric orbifold. We could then write the total
number of states schematically as
\be\label{allconj}
\rho(\Delta) \sim F(\Delta) e^{2\pi\Delta}\ ,
\ee
where $F(\Delta)$ is roughly the number of twisted
sectors that can contribute states of weight $\Delta$.
In particular (\ref{allconj}) would imply that
the growth behavior is mainly fixed by the number
or conjugacy classes of $G_N$: if they grow more slowly
than exponentially, then the phase diagram is the same
as for the symmetric orbifold. If they grow more quickly,
then that would change the phase diagram.

As a first step towards establishing (\ref{allconj}),
let us investigate the second factor in (\ref{Bantay})
more carefully.
Fix a configuration $g$ with cycle lengths $L_i$.
Let us concentrate for the moment on the term
with the trivial centralizer element $h_2=1$.
The contribution is then
\be
Z(\tau) = \prod_i Z(\frac{\tau}{L_i})\ .
\ee
The number of states of weight $\Delta$ coming from states of
weight $\Delta_i$ from the $i$th cycle, $\Delta= \sum_i \Delta_i$, is
\be
\rho_{G_N}(\Delta) = \prod_i \rho\left(L_i(\Delta_i-L_i\frac{c}{12})+\frac{c}{12}\right)\ ,
\ee
which of course vanishes unless for all $\Delta_i$
\be
\Delta_i \geq \frac{c}{12}(L_i-\frac{1}{L_i})\ .
\ee
Assuming for the moment that we are in the Cardy regime
for all the factors so that
\be\label{Cardy}
\rho \sim \exp\left(2\pi \sqrt{cL_i(\Delta_i-L_i\frac{c}{12})/3}\right)\ ,
\ee
we can evaluate the total contribution coming from all partitions
$\Delta_i$ by saddle point approximation.
The saddle point is given by
\be
\Delta_i = L_i(\frac{c}{12}+\lambda)\ , 
\ee
where the Lagrange multiplier is fixed by $\lambda = \Delta/L - c/12$
where $L = \sum_i L_i$,
so that in total the contribution of this configuration is
\be
\exp( 2\pi \sqrt{cN/3(\Delta-cN/12)})\ .
\ee
This suggests that the maximum does not even depend
on the specifics of the cycle decomposition, but only
on its total length. It is maximized for $L = \frac{6\Delta}{c}$ giving
indeed
\be\label{Hagedorn}
e^{2\pi \Delta}\ .
\ee
The issue is that we need to be more careful about the validity
of applying the Cardy formula (\ref{Cardy}).
Using (\ref{Cardy}) is valid only if
\be\label{conforCardy}
c \ll L_i\left(\Delta_i-L_i\frac{c}{12}\right) = L_i^2\lambda = L_i^2 \frac{c}{12}\ ,
\ee
\ie $L_i \gg 1$ for all $L_i$. This strengthens the result
in \cite{Belin:2014fna}: To ensure (at least) Hagedorn growth at $\Delta$,
it is sufficient to have an element in $G_N$ which has
several cycles $L_i\gg1$ such that $\sum_i L_i = L = 6\Delta/c$. The cycles
$L_i$ themselves can be much shorter than $L$.

Note however that there are two major caveats here. First of all,
the behavior for a $g$ which consists of many short cycles $L_i$
can be quite different, since in that case (\ref{conforCardy}) may
be violated, so that the Cardy formula may not apply. The main
worry here is that this may lead to growth faster than (\ref{Hagedorn}).
For example we can take $g$ to be given by $n$ cycles of length $L=2$.
The contribution of states of weight $\Delta_i=c/6$ to 
the state of total weight $\Delta=cn/6$ is then
\be\label{L2n}
\rho_{G_N}(\Delta) = \rho(c/12)^{\frac{6\Delta}{c}}\ .
\ee
If we choose a seed theory with a large $\rho(c/12)$,
then this seems to imply that we get indeed a faster
growth than (\ref{Hagedorn}).

The reason that this probably does not happen is related to
the second caveat. We have so far only considered the term $h=1$,
that is, we have not projected to $G_N$ invariant states.
We expect that this projection will eliminate most of
the states in (\ref{L2n}), since for a $g$ with so many
short cycles, the centralizer group is very large. On the other
hand for $g$ consisting of only a few long cycles,  
$C_g$ will be relatively small, and we expect (\ref{Hagedorn})
to hold to good approximation even after projecting to
invariant states.

This makes it plausible that something like (\ref{allconj}) could
indeed be true. To prove it however clearly more work is needed.

\subsection{Ramond and Neveu-Schwarz sectors}
From our arguments it is clear that it is crucial
that there is a vacuum in the theory, and that
it is non-degenerate. For purely bosonic theories this is never an issue,
since this is guaranteed by cluster decomposition.
In theories with fermions we have to be somewhat more careful.
For such theories this is still 
the case in the NS sector. In the Ramond sector however
the situation is more complicated, since the ground state
is no longer the vacuum, and therefore is no longer necessarily
non-degenerate.
In principle one can repeat the
same analysis also in the Ramond sector, or possibly in some mixed
NS-R sector. For $N=2$ theories it is usually assumed that
the results should be equivalent due to the spectral flow
symmetry of the theory. This is however only the case
if one keeps track of the $U(1)$ charges. Once one specializes
to different fugacities, there is no longer a guarantee that
the results will agree. This is especially severe in
the case at hand because spectral flow morally speaking
shuffles states around by a distance $c$, so that in the
large $c$ limit the process becomes even more drastic.

As an example for this phenomenon take for instance
the symmetric orbifold of $K3$. In the NS-NS sector,
the number of states remains of course perfectly finite,
as $S_N$ is oligomorphic. The free energy moreover is
the contribution of the vacuum with at most finite
corrections \cite{Keller:2011xi}. For the NS-R sector,
the situation is different. The lowest lying state
has degeneracy $N$ coming from the $N$ right-moving
Ramond ground states. This then leads to a logarithmic
correction to the free energy.
For other permutation orbifolds, the difference may
be even bigger. In \cite{Benjamin:2015hsa} the number of ground states
for the $S_{\sqrt{N}}\wr S_{\sqrt{N}}$ orbifold
of the $K3$ theory in the NS-R sector was found to be
$\sim e^{\sqrt{N}}$, which implies a correction
of order $O(\sqrt{N})$ to the free energy, which
is more than logarithmic corrections expected
from supergravity.
Note that all these states come from the untwisted
sector. We on the other hand have found in (\ref{wreathUntwisted}) 
that the 
growth of untwisted states in the NS-NS sector is perfectly
sub-Hagedorn, so that there is at most a finite $O(1)$
correction to the vacuum contribution to the free
energy. Although it is theoretically conceivable
that the twisted states could change the behavior,
the criterion given in \cite{Benjamin:2015hsa}
for the number of ground states is most likely not
a necessary condition for having a Hawking-Page transition
in the NS-NS sector.

\section{Factorization for the Symmetric Orbifold}
\subsection{General setup}
Let us now turn to correlation functions of permutation orbifolds.
The computation of correlation functions is in general much harder
than the counting of states done in the previous section.
One technique is to go to the cover of the underlying correlations
\cite{Dixon:1986qv}.
For symmetric orbifolds some correlation functions were indeed
evaluated in \cite{Lunin:2000yv,Lunin:2001pw}. In the large $N$
limit, \cite{Pakman:2009zz,Pakman:2009ab,Pakman:2009mi} argued that the cover method
leads to a diagrammatic $1/N$ expansion. Luckily for us,
we will not need such sophisticated methods. In fact 
we will argue that the leading contribution is always
very simple for the permutation groups in question,
and does not depend on the dynamics of the underlying
seed theory.

Let us now discuss factorization of the correlation function in the large
$N$ limit. Factorization means that any correlation function can be written
as the sum of products of two point functions. To put it another
way, any correlation function can be evaluated using Wick contractions,
so that the theory is a generalized free field. From the gravity
side we do indeed expect holographic CFTs to 
satisfy this property in the large $N$ limit --- see
\eg \cite{ElShowk:2011ag}. Ultimately we want to understand what
conditions this imposes on the permutation groups $G_N$.
In this section, as a warm up we will prove that symmetric
orbifolds indeed factorize in the large $N$ limit. This
is of course expected, as famously they are
dual to the $D1-D5$ system. In fact factorization
for single cycle twist fields was already argued in \cite{Lunin:2000yv}.

As we have argued above, 
a general state $\Phi$ of length $K$  
is given by fixing a $K$-tuple and summing over all its images
under the action of $G_N$. To compute the correlation
function of $n$ properly normalized fields,
we thus evaluate a total of $|G_N|^n$ terms.
A 2-point function has $|G_N|^2$ terms, and we will use it
to fix the norm of $\Phi$ to 1. A 3-point function then
has $|G_N|^3$ terms, so naively it seems like it
should diverge as $N\rightarrow\infty$. It turns out
however that a great many of those terms vanish, so that
(at least in the cases we discuss below) the result remains
finite.

In what follows it will be convenient to work with (unordered)
$K$-sets $\K$ of distinct elements rather than with ordered $K$-tuples $\vec{K}$.
As usual the two give the same result up to $N$-independent
factors. It is useful to present such a set $\K$ pictorially
as a row of black and white dots, \eg represent
\be\label{rowexample}
\K=\{1,2,5,7,8,9\} \qquad \textrm{as} \qquad \overbrace{\bb\bb\wb\wb\bb\wb\bb\bb\bb\wb\wb\wb\cdots\wb}^N\ .
\ee
In terms of the tensor product state, a white dot
thus corresponds to a \emph{trivial factor}, \ie
the corresponding factor is untwisted
and in the vacuum. A black dots denotes
a \emph{non-trivial factor}, which means
it is either untwisted, but not in the vacuum,
or it is part of a twist cycle.

When computing a 3-point function, each term corresponds
to three rows of the form (\ref{rowexample}) lined up
below each other. Each of the $N$ columns then contains
$i=0,1,2$ or $3$ non-trivial factors. Let us denote
the number of columns with $i$ non-trivial factors by $n_i$.
Note that if $n_1>0$, then this term directly vanishes: if the non-trivial
factor is untwisted, then it leads to a non-trivial 1-point function
in the underlying seed theory, which vanishes. If it is
part of a twisted cycle, then we know that the correlators
vanishes unless the twist sectors $g_i$ of the states involved
satisfy
\be
g_1g_2g_3 =1 \ .
\ee
This is clearly impossible if only one state has a twist
cycle in a given factor.

Let us now state the result for the symmetric orbifold:
For properly normalized states $\Phi_i$,
the total contribution of terms with total triple overlap 
$n_3$ to the sum in the 3-point function goes like
\be\label{factorizeresult}
\sim O(N^{-n_3/2})\ 
\ee
for large $N$.
This shows in particular that the only contributions
that survive the large $N$ limit have $n_3=0$, so
that all factors come as double overlaps, \ie as
2-point functions. 
The theory thus indeed becomes free in the sense
that all three point functions can be obtained
by Wick contractions. In gauge theory language
we can identify non-trivial cycles and non-trivial
factors with single trace operators.

In the rest of this section
we will show (\ref{factorizeresult}) for the symmetric orbifolds,
and in section~4 we will discuss under what condition it also holds
for other permutation orbifolds.

\subsection{Untwisted correlation functions}
As usual we will start with untwisted states.
In our notation, a prestate $\phi$ is given
by $\vec{K}$ and 
\be
\vec{\varphi}= (\underbrace{\varphi_1,\ldots,\varphi_1}_{k_1},\ldots,
\underbrace{\varphi_I,\ldots,\varphi_I}_{k_I})
\ee
with $\sum_{i=1}^I k_i = K$. 
The actual state is then obtained as a sum over $S_N$,
\be
\Phi = (A_\Phi)^{-1/2} \sum_{g\in S_N} \phi_{(g.\vec{K},\vec{\varphi})}\ .
\ee
We first need to fix the normalization $A_\Phi$ by computing
the 2-point function 
\be
\langle\Phi|\Phi\rangle = (A_\Phi)^{-1} \sum_{g_1,g_2\in S_N} 
\langle\phi_{(g_1.\vec{K},\vec{\varphi})}|\phi_{(g_2.\vec{K},\vec{\varphi})}\rangle\ .
\ee
Clearly we can simply pull out the action of one of the $S_N$.
We use it to fix the non-trivial factors of the first state to lie
in the first $K$ factors, and obtain
an overall factor of $|S_N|=N!$. Evaluating the terms coming
from the permutations of the second $S_N$, it is clear that they
vanish unless they are of the form
\begin{eqnarray*}
\phi_1:&&\overbrace{\bb\bb\bb\bb\bb\bb\bb\bb\bb\wb\cdots\wb}^N\\
\phi_2:&&\underbrace{\bb\bb\bb\bb\bb\bb\bb\bb\bb}_{K}\wb\cdots\wb
\end{eqnarray*}
If we take the $\varphi_i$ to be orthonormal,
we can evaluate the sum over $S_N$ to fix the normalization constant as
\be\label{APhi}
A_\Phi= N! (N-K)!\prod_i k_i! \sim N!(N-K)! \ .
\ee
In the last expression, $N!$ comes from the order of the group $S_N$,
and $(N-K)!$ is the order of the pointwise stabilizer of the  
first $K$ elements. In what follows, we will often drop $N$-independent contributions if convenient, and 
use $\sim$ to denote equality up to $N$-independent factors in the equations,
just like we did in (\ref{APhi}).

Now we turn to three point functions.
Take three states $\Phi_i$ of length $K_1,K_2,K_3$. 
%
Consider all terms in the sum
with fixed $n_3$.
Schematically, they look like
\begin{eqnarray*}
\phi_1:&&\overbrace{\underbrace{\bb\bb\bb\bb\bb\bb\bb\bb\bb\bb\bb\bb\bb\bb}_{K_1}\underbrace{\wb\wb\wb\wb}_{K_2-J}\wb\cdots\wb}^N\\
\phi_2:&&\underbrace{\bb\bb\bb\bb\bb\bb\bb\bb\bb\bb}_J\wb\wb\wb\wb\underbrace{\bb\bb\bb\bb}_{K_2-J}\wb\cdots\wb\\
\phi_3:&&\underbrace{\bb\bb\bb\bb\bb}_{n_3}\wb\wb\wb\wb\wb\underbrace{\bb\bb\bb\bb\bb\bb\bb\bb}_{K_1+K_2-2J}\wb\cdots\wb
\end{eqnarray*}
For notational convenience we have used the overall $S_N$ to make the
non-trivial factors occupy only the first $K_1+K_2-J$
columns; for the argument that follows this does not matter. Also note
that for the same reasons as above $n_1=0$, so that this
is the only non-vanishing type of contribution.
$J$ is fixed by
\be
K_3 = K_1+K_2 -2J+n_3\ .
\ee
Let us now count the number of such terms in the sum over the
three symmetric groups $S^{(1)}_N,S^{(2)}_N,S^{(3)}_N$.
As usual we fix the overall $S_N$ and
pull out a factor $N!$ from $S^{(1)}_N$. Next consider $S^{(2)}_N$. Here we first distribute the $K_2-J$ factors
of $\phi_2$ over $N-K_1$ slots for which there are no $\phi_1$
states, giving $(N-K_1)!/(N-K_1-K_2+J)!$ possibilities, and then distribute the trivial
factors over the remaining slots, for which there are $(N-K_2)!$
possibilities.
Finally, for $\Phi_3$, the positions of the non-trivial factors are fixed
by the condition that they have to pair up with the remaining
non-trivial factors of $\Phi_{1,2}$, which leads to an $N$-independent
combinatorial factor only. The trivial states however can again be distributed
in $(N-K_3)!$ different ways. In total there are thus
\be\label{untwistedcentralizer}
\frac{N!(N-K_1)!(N-K_2)!(N-K_3)!}{(N-\frac{1}{2}(K_1+K_2+K_3-n_3))!}
\ee
terms.
Combining this with the normalization factor $(A_1A_2A_3)^{-1/2}$, we get
\be
\sim \left(\frac{(N-K_1)!(N-K_2)!(N-K_3)!}{N!((N-\frac{1}{2}(K_1+K_2+K_3-n_3))!)^2}\right)^{1/2} \sim O(N^{-n_3/2})\ ,
\ee
which indeed establishes (\ref{factorizeresult}).

As a special case, take $K_3=1$ with the only non-trivial state
given by the energy-stress tensor $T$. Naively our results imply
that the three point function should vanish as $N^{-1/2}$, which 
seems like a contradiction. Note however that
in the above derivation
we have chosen the states to be orthonormal, whereas the
correct normalization for $T$ is $\langle T|T\rangle\sim c\sim N$.
Using this normalization we need to multiply by $N^{1/2}$ and
so do indeed get that the 3pt function is finite,
\be
\langle OOT\rangle \sim O(1)\ ,
\ee
which is consistent with the usual OPE of the energy-stress tensor.

\subsection{Twisted sector}
Let us now discuss twisted states.
A twisted sector is given by a conjugacy class $[g]$ of $S_N$.
When computing correlation functions of states,
we first need to average over all elements of the conjugacy
class by picking a specific instance $g\in[g]$ and
then sum $hgh^{-1}$ over $h\in S_N/C_g$. 
We also need to project to an orbifold invariant state.
This means we need to sum over the centralizer $C_g$, \ie all elements
$h$ which commute with $g$. As we argued before, the combined
action of these two on $\vec{K}$ is just the standard action of the full
group $S_N$ on $\vec{K}$. Since the argument for factorization
only depended on counting the number of configurations,
it is clear that the essentially the same argument will
go through also for twisted states.

Let us start with the case where all non-trivial factors 
are twisted. For distinction we will denote
these factors by $\times$. The $N$-dependent
part of the action of the centralizer $C_g$
then simply factorizes through and cancels
with the norm, so that we can simply take
$S_N$ to act in the standard way.
The normalization of $\Phi$ is then again given by
\be
A_\Phi \sim N!(N-K)!
\ee
%
%
The three point function works out exactly
like in the untwisted case:
Consider all terms with $n_3$
triple overlaps of twisted factors,
\begin{eqnarray*}
\phi_1:&&\underbrace{\tw\tw\tw\tw\tw\tw\tw\tw\tw\tw\tw\tw\tw\tw}_{T_1}\underbrace{\wb\wb\wb\wb}_{T_2-J}\wb\cdots\wb\\
\phi_2:&&\underbrace{\tw\tw\tw\tw\tw\tw\tw\tw\tw\tw}_J\wb\wb\wb\wb\underbrace{\tw\tw\tw\tw}_{T_2-J}\wb\cdots\wb\\
\phi_3:&&\underbrace{\tw\tw\tw\tw\tw}_{n_3}\wb\wb\wb\wb\wb\underbrace{\tw\tw\tw\tw\tw\tw\tw\tw}_{T_1+T_2-2J}\wb\cdots\wb
\end{eqnarray*}
$S_N$ acts by conjugation on the $g_i$, which in this notation is
the same as its action on the factors in the untwisted case.
The counting of the terms is thus exactly the same as in
the untwisted sector, so that
in total we again obtain
\be\label{twistedconjugation}
\sim O(N^{-n_3/2})\ .
\ee
For completeness we want to argue that the case of $n_3=0$ does indeed reduce
to a product of two point functions.
Here it is important that $n$-point functions
vanish unless
\be\label{twistrules}
g_1g_2\cdots g_n =1 \ .
\ee
 Because $n_3=0$, the set of non-trivial factors $I$ decomposes into
\be
I = I_1 \sqcup I_2 \sqcup I_3\ ,
\ee
where $I_i$ is the subset of $I$ pointwise invariant under $g_i$.
$g_1$ then has to map $I_2$ and $I_3$ to themselves, \ie that $g_1 \in S_{I_2}\times S_{I_3}$:
If $g_1$ mapped an element $i_3\in I_3$ to $i_2\in I_2$,
then because $g_3g_2g_1=1$ and $g_2$ leaves $I_2$ invariant,
$g_3$ would have to map $i_2$ to $i_3$, which contradicts
that $I_3$ is invariant under $g_3$.
This and similar arguments for $g_{2,3}$ show that the
3pt function can indeed be evaluated as a product
of 2pt functions of twist fields.

The most general setup is a combination of twisted factors
and non-trivial untwisted factors like  
\begin{eqnarray*}
\phi_1:&&\tw\tw\tw\tw\tw\tw\tw\tw\tw\tw\bb\wb\wb\wb\bb\bb\bb\bb\wb\wb\cdots\wb\\
\phi_2:&&\tw\tw\tw\tw\tw\tw\tw\tw\wb\wb\tw\tw\tw\tw\bb\bb\bb\wb\bb\wb\cdots\wb\\
\phi_3:&&\tw\tw\tw\bb\bb\wb\wb\wb\tw\tw\tw\tw\tw\tw\bb\bb\wb\bb\bb\wb\cdots\wb
\end{eqnarray*}
From the discussion in section~2 we know that 
the conjugation and the centralizer act in the same
way on $\vec{K}$. The exact same counting argument
as above thus goes through, which gives
\be
\sim O(N^{-n_3/2})\ .
\ee

\subsection{Higher point functions}
In fact this Wick factorization property 
carries over to higher point functions.
To see this, decompose the correlation function
into 3pt functions and sum over
intermediate states.
The important thing to note is that for two states
of length $K_{1,2}$, all 3pt functions with
$K_3 > K_1+K_2$ vanish. This means that only 
states with a fixed, $N$ independent length
run in the intermediate channels,
even though of course for a given length
there are still an infinite number of states
which a priori give a non-vanishing contribution.
The contributions for a given length,
\ie the infinite sum over all states of such length, 
can however be computed
in a finite symmetric orbifold. This means
that they are finite, and 
do not depend on $N$, which shows that all 
the $N$ dependence comes from the 3pt functions.
It follows that the factorization arguments carry over
to general $n$-point functions.


\section{Factorization for general permutation groups}
\subsection{Factorization for oligomorphic groups}
Let us now discuss factorization for general oligomorphic group.
First we will rewrite much of the above in more group
theoretic language.
Let $\stab{K}$ be the subgroup of $G_N$ stabilizing the set $\K$.
Through the orbit-stabilizer theorem we can always relate this
to the length of the orbit of $\K$, $O_N(\K)$: 
\be
|\stab{K}| = |G_N|/O_N(\K)\ .
\ee
In particular note that $|\stab{K}|$ is independent
of which element of the orbit we choose.
Using this new notation the normalization factor of a state $\Phi$ comes out to
\be
A_\Phi \sim |G_N||\stab{K}|= \frac{|G_N|^2}{O_N({\K})}\ .
\ee
The formula for the 3pt function can be obtained in a similar fashion following the procedure of $S_N$. 
Consider the again configuration
\begin{eqnarray*}
\phi_1:&&\overbrace{\underbrace{\bb\bb\bb\bb\bb\bb\bb\bb\bb\bb\bb\bb\bb\bb}_{K_1}\underbrace{\wb\wb\wb\wb}_{K_2-J}\wb\cdots\wb}^N\\
\phi_2:&&\underbrace{\bb\bb\bb\bb\bb\bb\bb\bb\bb\bb}_J\wb\wb\wb\wb\underbrace{\bb\bb\bb\bb}_{K_2-J}\wb\cdots\wb\\
\phi_3:&&\underbrace{\bb\bb\bb\bb\bb}_{n_3}\wb\wb\wb\wb\wb\underbrace{\bb\bb\bb\bb\bb\bb\bb\bb}_{K_1+K_2-2J}\wb\cdots\wb
\end{eqnarray*}
where
\be
J = \frac{1}{2}(K_1+K_2-K_3+n_3)\ .
\ee
Note that this is only a very schematic picture of the situation: 
For general $G_N$ there is certainly no guarantee that we can move
all the non-trivial factors all the way to the left. The 
position of the columns
should therefore be understood up to permutation. 
We need to estimate the number of such terms.
The sum over $G_N^1$ gives again $|G_N|$. The sum over $G_N^2$ is more subtle. We will get $|\stab{K_2}|$ for the vacuum states, but we must also sum over the different ways the non-trivial factors of $\phi_2$ can distribute themselves over the vacuum states of $\phi_1$.
This number is given by the stabilizer of $\K_1$ modulo the stabilizer of $\K_1\cup\K_2 = \K_1\cup\K_2\cup\K_3$,
\be
|\stab{K_1}|/|G_N^{\K_1\cup\K_2\cup\K_3}|\ .
\ee
The trivial factors of $\phi_3$ again give $|\stab{K_3}|$. Including the normalization factors,
the total contribution is thus
\be\label{3pt}
\frac{(O_N({\K_1})O_N({\K_2})O_N({\K_3}))^{1/2}}{|G_N|^3} \frac{|G_N| |\stab{K_1}| |\stab{K_2}||\stab{K_3}|}{|G_N^{\K_1\cup\K_2\cup\K_3}|}
= \frac{O_N(\K_1\cup\K_2\cup\K_3)}{(O_N({\K_1})O_N({\K_2})O_N({\K_3}))^{1/2}}\ .
\ee
Let us now make an additional assumption. We will call
$G_N$ \emph{democratic} if for fixed $K$, all orbits
have the same length up to $N$ independent factors.
Up to factors the orbit length is thus only a function of $K$,
so that we can write (\ref{3pt}) as
\be\label{3ptdemocratic}
\sim \frac{O_N(\frac{1}{2}(K_1+K_2+K_3-n_3))}{(O_N({K_1})O_N({K_2})O_N({K_3}))^{1/2}}\ .
\ee
If $G_N$ is oligomorphic, we can estimate the average 
orbit length as
\be
\langle O_N({\K}) \rangle = \binom{N}{K} f_K^{-1} \sim \binom{N}{ K}\ 
\ee
where have used that $f_K$ is independent of $N$
for $N$ large enough.
Democracy then implies that all orbits have average length,
\be\label{democratic}
O_N({\K}) \sim \langle O_N({\K}) \rangle \sim \binom{N}{ K} \qquad \forall |\K| = K\ .
\ee
Plugging this into (\ref{3ptdemocratic}) we recover the same result
as for symmetric orbifolds,
\be
\sim N^{\frac{1}{2}(K_1+K_2+K_3-n_3) - \frac{1}{2}(K_2+K_3+K_1)} \sim N^{-n_3/2}\ .
\ee
This shows that democratic oligomorphic groups factorize in the large $N$ limit.

\subsection{Relaxing the assumptions}
In the previous section we showed that democratic and oligomorphic
are sufficient for factorization. We believe that those two
conditions are too strong, and can be relaxed significantly.
Let us discuss a few examples to support this belief.

First consider the cyclic orbifold.
Untwisted states are given by $N$-tuples of
states up to cyclic shifts. In this case
we define the length $K$ as the number of non-trivial
factors, which again has to remain finite for the
state to have finite weight. Since $\Z_N$ is Abelian,
the twisted sectors are simply given by elements of $\Z_N$,
which is given by $N_n$ cycles of length $n$ such that $n N_n=N$.
The weight of the ground state of this sector is given
by 
\be
\Delta = \frac{c}{24}\sum_i (n -\frac{1}{n}) = \frac{c}{24}(N-N/n^2)
\ee
which shows that only the untwisted sector with $n=1$
has finite large $N$ limit. In what follows we can
thus concentrate on the untwisted sector only.

In the untwisted sector, we have $O_N(\vec{K}) = N$
for all $K >0$, similarly for $O_N(\K)$. The group is thus
clearly democratic. Applying this to (\ref{3ptdemocratic}),
we find that the correlator of three non-trivial fields
goes like
\be
\sim N^{-1/2}\ .
\ee
The only exception is if one of the states is
the vacuum, in which case we are back at a two point function
which of course goes as $\sim 1$.
Cyclic theories thus also become free in the large $N$
limit, albeit in a somewhat trivial way. They do have an infinite number of
states at fixed energy. This shows that there
are non-oligomorphic orbifolds which factorize.

What happens if we relax the democratic assumption?
It is then possible that different of orbits of $K$-tuples
have vastly different lengths. An example of this is
\be
G_N = \mathbf{1} \times S_{N-1}\ .
\ee 
The 1-tuples have two orbits, one of length 1,
the other of length $N-1$. Since this choice
of $G_N$ gives a symmetric orbifold with
an additional tensor factor, it is clear that
the 3pt functions do not factorize, since the first
factor does not.

Still, it seems very likely that also many
non-democratic groups factorize. One guess would
be that permutation groups factorize if they are \emph{transitive},
\ie if their natural action is transitive.
This has the added advantage that they will
have a unique energy-momentum tensor.

\subsection{$S_{\sqrt{N}}\wr S_{\sqrt{N}}$}
Let us now consider $S_{\sqrt{N}}\wr S_{\sqrt{N}}$.
As we argued above, instances of orbits
are given by partitions of $K$. For a given partition $P_K$, $\{k^l\},l=1\ldots L$, $\sum_l k^l = K$,
the stabilizer is given by
\be
|\stab{K}| \sim  |S_{\sqrt{N}-L}| |S_{\sqrt{N}}|^{\sqrt{N}-L} |\prod_l |S_{\sqrt{N}-k^l}| \sim N^{\frac{1}{2}(\sqrt{N}-L+N-\sqrt{N}L +\sqrt{N}L-K)}
\ee
so that the orbit length is given by
\be
O(\K) \sim N^{\frac{1}{2}(K+L)}\ .
\ee
This shows that the wreath product is not democratic,
since the length of the orbit not only depends on $K$,
but also on the number of columns $L$ of the partition.
A quick and dirty argument
then shows that wreath product factorizes:
For the configuration $\K_t= \K_1\cup\K_2\cup\K_3$ we have 
$K_t = \frac{1}{2}(K_1+K_2+K_3-n_3)$ and $L_t = \frac{1}{2}(L_1+L_2+L_3-n^c_3)$
where $n^c_3$ is the number of triple overlaps
of non-trivial rows.
From (\ref{3pt}) we thus get in total
\be
\sim O(N^{-\frac{1}{4}(n_3+n^c_3)})\ .
\ee
Let us give a more careful argument for this using
a detailed counting of the terms.
Fix $\K_1$ to be arranged as a partition
of $K_1$ in the upper left corner, pulling out an overall
factor of $|\wrS|$. Next we take an image of $\K_2$ which has 
$k_2^l$ non-trivial factors in the $l$-th row with $l=1,\ldots\sqrt{N}$. 
For $N$ large of course most rows will have $k_2^l=0$.
Consider configurations where there
are $J$ non-trivial rows of $\K_2$ in the first $L_1$ lines.
There are $\sim \binom{\sqrt{N}-L_1}{L_2-J}$ such configurations,
each of which comes with an additional factor $|S_{\sqrt{N}-L_2}|$
from distributing the trivial rows. Together this gives
\be
\sim N^{\frac{1}{2}(\sqrt{N}-J)}\ 
\ee
ways of how the non-trivial rows of $\K_2$ can be distributed.
Next let us count for each row in how many ways the factors can be
distributed.
Take a configuration for which of the $k_2^l$ non-trivial factors in
row $l$, $J^l$ overlap with non-trivial factors of $\K_1$.
For the $l$-th row we then have $\sim \binom{\sqrt{N}-n^l_1}{n^l_2-J^l}$
possible ways to distribute them, with an additional factor of 
$|S_{\sqrt{N}-n_2^l}|$ from the trivial factor. Taking the product
over all rows we get
\be
\sim \prod_{l=1}^{\sqrt{N}} N^{\frac{1}{2}(\sqrt{N}-J_l)} = N^{\frac{1}{2}(N-J_{tot})}\ ,
\ee
where $J_{tot}$ is the total number of $\K_1$ and $\K_2$ factors
that overlap.
Since 1pt functions vanish, the distribution of non-trivial $\K_3$ factors is then again
completely fixed up to $N$-independent factors, so that
that we get an overall contribution of the stabilizer $|\stab{K_3}|$.
Putting this together with the normalization we get
\begin{multline}
\frac{(O_N({\K_1})O_N({\K_2})O_N({\K_3}))^{1/2}}{|G_N|^3} |G_N| |\stab{K_3}| N^{\frac{1}{2}(N+\sqrt{N}-J-J_{tot})}\\
=\frac{(O_N({\K_1})O_N({\K_2}))^{1/2}}{O_N({\K_3})^{1/2}|G_N|} N^{\frac{1}{2}(N+\sqrt{N}-J-J_{tot})}\\
= N^{\frac{1}{4}(K_1+K_2-K_3 - 2J_{tot} +L_1+L_2-L_3 - 2J)} = N^{-\frac{1}{4}(n_3 + n^{c}_{3})}\ ,
\end{multline}
where the $n^{c}_3$ is the number of non-trivial row triple overlaps.
This shows that the wreath product indeed factorizes.

\bigskip

\textbf{Acknowledgments:} We thank Nathan Benjamin, Alejandra Castro, Miranda Cheng, Ethan Dyer, Liam Fitzpatrick, Felix Haehl, Shamit Kachru, Josh Lapan, Arnaud Lepage-Jutier, Greg Moore, Mukund Rangamani and Shlomo Razamat for useful
discussions. CAK thanks the Perimeter Institute, the Harvard University High Energy Theory Group
and the Aspen Center for Physics for hospitality,
where part of this work was completed. CAK was supported by the Rutgers
New High Energy Theory Center, U.S. DOE Grants No.~DOE-SC0010008,
DOE-ARRA-SC0003883 and DOE-DE-SC0007897 and in part by National Science Foundation Grant No.~PHYS-1066293. AB
is supported by the Swiss National Science Foundation, Grant No. P2SKP2\textunderscore158696.
AM is supported by the National Science and Engineering Council of Canada.

\bibliographystyle{ytphys}
\bibliography{ref}

\end{document}